\documentclass[a4paper]{article}
\usepackage[ansinew]{inputenc}
\usepackage{times}
\usepackage[T1]{fontenc}
\usepackage{graphicx}
\graphicspath{{./Graphiken/}}
\usepackage{booktabs} 
\usepackage{tabularx} 

\usepackage{geometry}
\usepackage{amssymb}
\usepackage{amsmath}

\usepackage{rotating}
\usepackage{tcolorbox}
\usepackage{xcolor}
\colorlet{shadecolor}{gray!25}
\usepackage{subfig}
\usepackage{float}

\author{\normalsize Ludwig A. Hothorn,\\ 
\footnotesize Im Grund 12, D-31867 Lauenau, Germany (e-mail:ludwig@hothorn.de)\\ \scriptsize(retired from Leibniz University Hannover)}

\title {Consistent ANOVA-type tests for various effect sizes}

\begin{document}
\maketitle

\section{Introduction}\label{sec1}

The analysis of variance (ANOVA) is widely used in practice and provides excellent properties from a statistical perspective. However, it includes some disadvantages, such as non-robustness against heteroscedastic and/or non-normal errors,  or as providing global decision by means of a single p-values only. A disadvantage, which has been little discussed so far, is its exclusive formulation for mean differences from the overall mean as effect size. Some applications require alternative effect sizes for problem-appropriate interpretation. Moreover, in some studies or trials multiple primary endpoints have to be consider and they may have quite different scales. The standard ANOVA is difficult to modify for this purposes. Without significant limitation of generalizability, only an one-way layout is considered here, just for simplicity.\\
The first basic property of the proposed approach is the similar power behavior of the ANOVA F-test and the multiple contrast test (MCT) comparing to the overall mean (OM) \cite{Konietschke2013}. The second basic property is the relative simple derivation of MCT's for different effect sizes \cite{Hothorn2008}. Here are specific considered: i) ratio-to-OM, ii) quantiles for both ratio or differences to OM, iii) odds ratios for continuous data,, iv) odds ratio for proportions, v) risk ratio and differences, vi) relative effect size for continuous up to discrete data, and vii) hazard ratio to OM.  The effect size is commonly determined by the scale of the endpoints, e.g. a proportion, by the chosen design and underlying randomization principle, e.g. completely randomized design, by the choice of interpretation, e.g. additive vs. multiplicative effect, and more issues.

\section{MCT against overall mean- the standard parametric case}\label{MCT}
MCT against overall mean provides either simultaneous confidence intervals or compatible p-values as well as a global p-value by means of the min-p approach \cite{Pallmann2016}. MCT is an union-intersection test $t_{MCT}=max(t_1,...,t_\xi)$ based on the single contrasts  $	t_{Contrast}=\sum_{i=0}^k c_i\bar{y}_i/S \sqrt{\sum_i^k c_i^2/n_i}$ where $c_i^q$ are the contrast coefficients. For MCT against OM the $c_i^q$ are simple (here for $i=3$ treatment groups $T_i$ in a balanced design, to keep it simple):

        \begin{table}[H]
				\centering
        \begin{tabular}{ c  c c r c c c c }
         $c_i$ & $T_1$ & $T_2$ & $T_3$ \\ \hline
        $c_a$ & -1 & 1/2   & 1/2   \\
        $c_b$ & 1/2 & -1 & 1/2  \\
				$c_b$ & 1/2 & 1/2 & -1  \\
        \end{tabular}
				\caption{Grand Mean comparisons contrasts} 
				\end{table}

The maxT test is a scalar test but its multiple test statistics $(t_1,...,t_\xi)'$ follows jointly a $\xi$-variate $t$-distribution with the common degree of freedom $df$ and correlation matrix  $R$ ($R=f(c_{ij},n_i)$) based on a common variance estimator $S$ and common $df$. Both two-sided hypotheses as well as one-sided hypotheses can be formulated. Size and/or power problems arise for designs with small $n_i$ because most of the proposed generalizations are asymptotic approaches only \cite{Hothorn2008}. \\
Extensions to factorial designs particularly for the interesting interactions are possible as long the number of factors and the number of factor levels is small (due to interpretation, not estimation or computation) \cite{hothorn2022IA}.

\subsection{A motivating example}\label{Exa1}
The library(gamlss) provides the raw data of the Hodges data \cite{hodges1998} containing (among others) the average monthly premium (in US dollars) (prind) for individual subscribers in state-based health maintenance organizations (HMO). Although the objective of the original paper is  complex hierarchical modeling, here the HMO's are simply considered as replicates within  only 41 US states (whereas $n_i=1$ data were omitted):

\begin{figure}[htbp]
	\centering
		\includegraphics[width=0.8\textwidth]{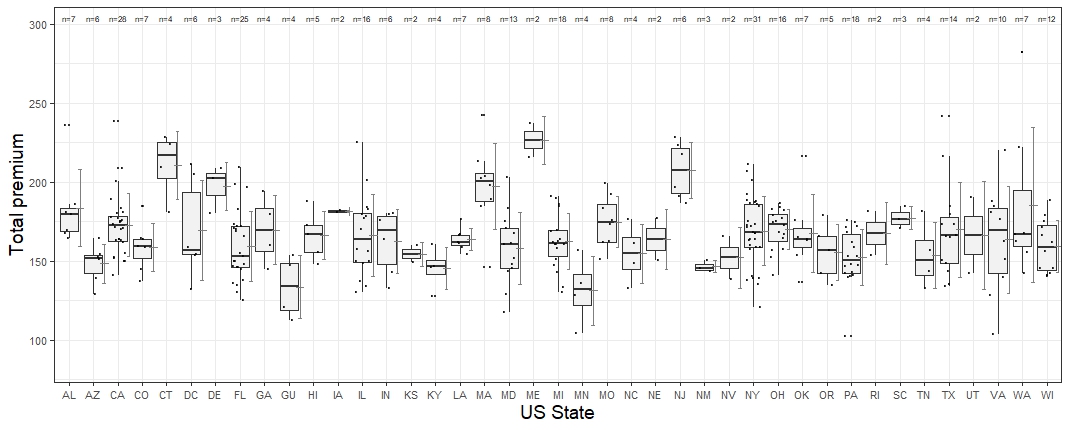}
	\caption{Box-plot of modified Hodges (1998) data}
	\label{fig:BoxHodges}
\end{figure}

This data example is highly unbalanced in an one-way layout with $k=41$ factor levels whereas various patterns of heteroscedasticity are obvious.

\section{MCT against overall mean- the modification for heteroscedastic errors}\label{hetMCT}

It is well known that with heterogeneous variances, especially with inverse unbalancedness, the test level is no longer maintained and the power is reduced. Therefore, appropriate modifications, e.g. with Welch-type df's or sandwich estimators (instead of the joint MQR) are available \cite{Herberich2010, Hasler2008}. However, the simultaneous confidence intervals (or the adjusted p-values) may be additionally distorted if, for example, a significantly higher variance occurs in a treatment group of no interest \cite{hothorn2023het}. Therefore, the Welch-type modification is proposed as the standard variant, especially for the small $n_i$ considered here.\\

The related R-Code is simple:
\small
\begin{verbatim}
library(multcomp)
library(sandwich)
mod1<-lm(prind~state, data=myhod)
plot(glht(mod1, linfct = mcp(state = "GrandMean"), vcov=vcovHC))
\end{verbatim}
\normalsize
The resulting simultaneous confidence intervals for the difference to OM are:
\begin{figure}[H]
	\centering
		\includegraphics[width=0.79\textwidth]{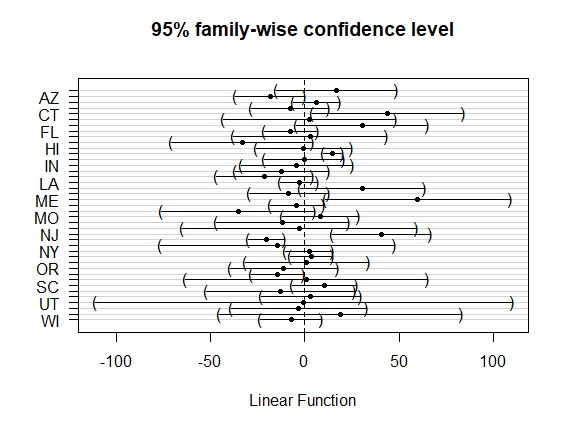}
	\caption{Simplified Hodges data: sCI for difference to OM }
	\label{fig:CIdiff}
\end{figure}
There are no increased rates compared to the overall mean, but reduced rates in NM, AZ, PA (in this order). The different widths of the sCI's are clearly visible, e.g. UT (std. error 33.8) and IA (std. error 1.6).

\section{MCT for ratios-to-overall mean}\label{ratio}
Even though the difference is used widely as effect size, ratios-to-OM as effect size in the multiplicative model show advantages, e.g. the dimensionless for easy interpretation of percent change, increase of power. For the hypotheses in the multiplicative model  
$H_{0i}: \frac{\mu_{i}}{\mu_{OM}}\leq\theta \quad {\rm{vs.}} \quad H_{1i}:  \frac{\mu_{i}}{\mu_{OM}}>\theta$ 
are the estimation of simultaneous Fieller-type confidence intervals challenging based  on the contrast test $ T_{i}=\frac{\bar{X}_{i}-\theta\bar{X}_{OM}}{S\sqrt{\frac{1}{n_{i}}+\frac{\theta^{2}}{n_{OM}}}}$ \cite{Dilba2004}
For the case of variance heterogeneity the related function \textit{sci.ratioVH()} is available in the CRAN package \textit{mratios} \cite{Dilba2007}
\small
\begin{verbatim}
library(mratios)
plot(sci.ratioVH(prind~state, data=myhod,type = "GrandMean"))
\end{verbatim}
\normalsize
The related simultaneous confidence limits reveal quite different pattern compared with difference-to-OM, i.e. increased rates compared to the overall mean in IA, NJ but reduced rates in NM and PA. From the boxplots in Figure \ref{fig:BoxHodges}, one would infer increases for ME,CT,NJ and decreases for MN, GU from the location of the medians. The confidence intervals show a different picture, not surprisingly given the extremely different $s_i$ and $n_i$. The finding for IA should be critically questioned, since it is largely due to the extremely low variance of the two values $(182.1; 180.8)$ with a global value range of about $(100-275)$.
\begin{figure}[htbp]
	\centering
		\includegraphics[width=0.92\textwidth]{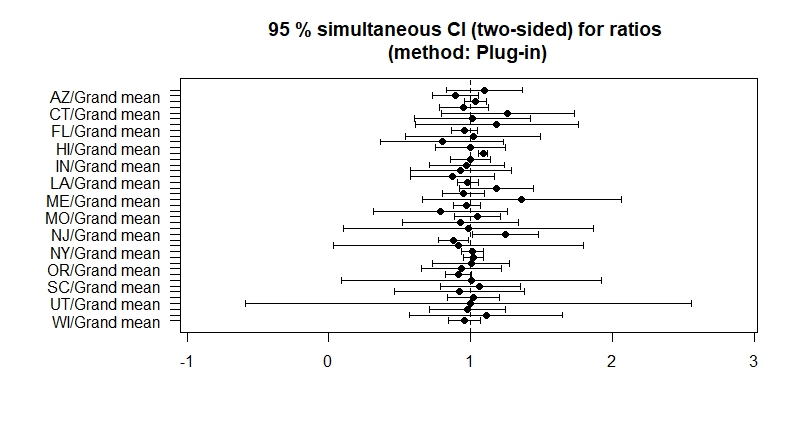}
	\caption{Modified Hodges data: simultaneous confidence intervals for ratio-to-OM}
	\label{fig:CIratioplot}
\end{figure}

The different widths of the sCI's are clearly visible,  whereas the degree of freedom varies from 2 to 20.1. 

\section{MCT for relative effect size}\label{konie}
In a nonparametric framework, the relative effect sizes can be defined as  specific proportions for two selected pairwise treatments 
$(\vartheta,\varsigma)$ to:  $p_{\vartheta,\varsigma}=Pr(X_{\vartheta} < X_{\varsigma})+\frac{1}{2} Pr(X_{\vartheta} = X_{\varsigma})$ 
(where $p_{\vartheta,\varsigma}=0$ at the null  hypothesis). Based on joint pseudo ranks the max T test $T_{max}=max(T_1,...,T_\xi)$ is for $l\in (1,...,\xi)$ multiple contrasts $T_l=\sqrt{N}\frac{\hat{p_l}-0.5}{\sqrt{\hat{\nu_l}}}$ with $p_l=\int{(\sum \left|c_{li} \right|F_i)d(\sum c_{lj}F_j)}$. The $(T_1,...,T_{\xi})'$ follows jointly an approximate $\xi$-variate normal distribution $z_{\xi,two-sided,1-\alpha,R}$ with correlation matrix $R$ whereas for small sample size an approximate t-distributed  version with related Welch-type $df^*$ \cite{Brunner2000} is recommended.  Related confidence intervals based on  $[0,1]$ range preventing transformations are available in the package \textit{nparcomp} \cite{Konietschke2015}. For the above Hodges data, the R-code is:
\small
\begin{verbatim}
library(nparcomp)
npc<-mctp(prind~state, data=myhod, type ="GrandMean", alternative ="two.sided",
          asy.method = "mult.t", info = FALSE,  correlation = FALSE)
\end{verbatim}
\normalsize
Non of the states are significantly different from OM. As effect size the log odds ratio can be used alternatively, whereas the width of the intervals in this example are huge.

\section{MCT for quantiles}\label{schaar}
Particularly for skewed distributions quantiles, particularly medians, are commonly used as effect size. Simultaneous confidence intervals for quantiles are recently available for differences or ratios \cite{segbehoe2022}. This approach can be used for comparisons against OM easily:
\small
\begin{verbatim}
library(mratios)
quantRatio<-mcpqrci(myhod$prind, myhod$state, p = 0.5, conf.level = 0.95, 
            type = "GrandMean", method =c("Fieller"), dist="MVT")
quantD<-mcpqdci(myhod$prind, myhod$state, p = 0.5, conf.level = 0.95, 
        type = "GrandMean", dist="MVT")
\end{verbatim}
\normalsize

The related simultaneous confidence intervals for the above Hodges data for ratio-to-OM and difference-to-OM are (whereas the states are listed in alphabetic order as C1,...,C41):

\begin{figure}[htbp]
	\centering
		\includegraphics[width=0.49\textwidth]{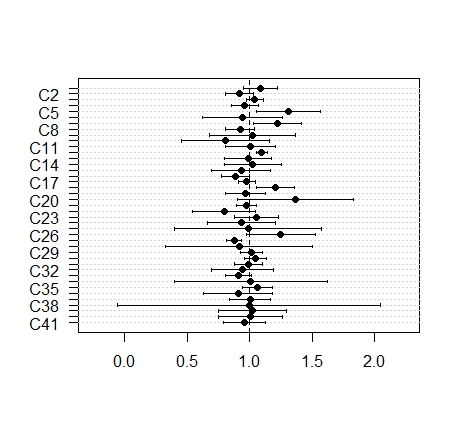}
		\includegraphics[width=0.49\textwidth]{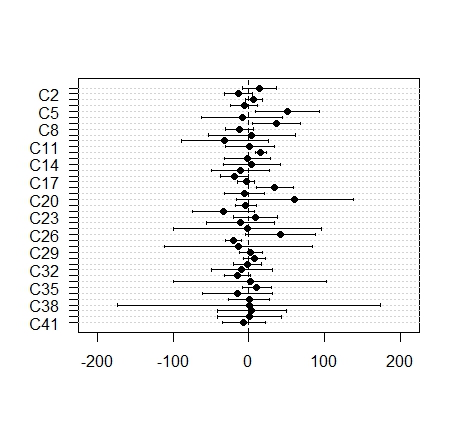}
	\caption{Hodges data: sCI for medians- ratio (left panel) and difference-to-OM (right panel)}
	\label{fig:Qratio}
\end{figure}
The pattern of significant changes are the same for both ratio and difference(increased CT, DE, IA, MA, decreased KY, NM) and the question arises what means the at least increase of $105.7\%$ (lower limit) for MA (C18 in the figures) in practice?

\section{MCT for odd ratios for continuous endpoints}\label{tram}
The basic paradigm to demonstrate the change between two distributions by means of the t-test, is to consider mean differences only (1st moments), assume homogeneous variances (2nd moment) (or adjust against variance heterogeneity in the Welch-t-test) and ignore the higher moments. Tests for distribution function differences, like Kolmogorov-Smirnov test, fail on the one hand because of the high requirements of $n_i$ and difficult interpretation with respect of effect size. A related, more recent concept represents the most likely transformation approach (MLT) \cite{Hothorn2018}, which is based on multiple transformation models and allows the most appropriate model to be selected within a maximum likelihood framework.
Whereas the common regression models estimate the conditional mean as a function of the  covariate (or factor) assuming variance or skewness can be ignored, the MLT approach takes these into account. Therefore, this approach is robust to non-normal distributions (including discrete ones), variance heterogeneity, and extreme values, such as data at detection limit(s)). Here, the variant based on continuous outcome logistic model approach \cite{Lohse2017} is demonstrated, where the resulting effect size log-odds ratio allows a scale-independent interpretation. The use with MCT's is possible \cite{Hothorn2019}. A related CRAN package \textit{mlt} is available \cite{Hothorn2020}.\\

The related R-code is not complicated whereas an asymptotic version was used:
\small
\begin{verbatim}
library(tram)
MLT<-glht(Colr(prind~state, data=myhod), linfct = mcp(state = "GrandMean"))
\end{verbatim}
\normalsize

\begin{figure}[htbp]
	\centering
		\includegraphics[width=0.79\textwidth]{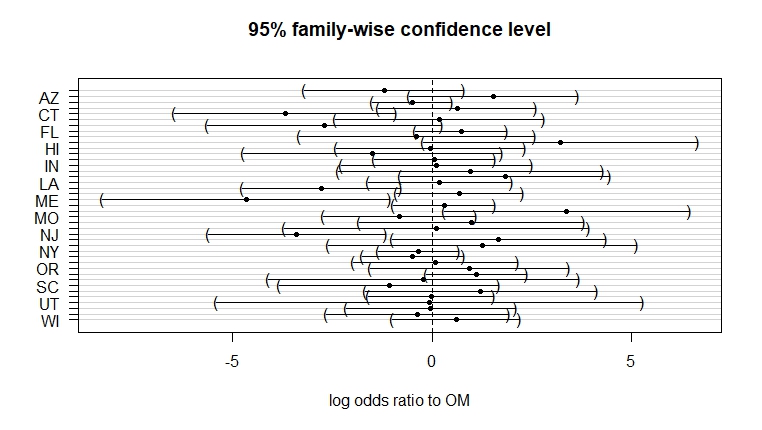}
	\caption{Modified Hodges data:  most likely transformation model}
	\label{fig:CImlt}
\end{figure}

The pattern is quite different from the above methods: CT,MA,ME, MN and NJ are significantly reduced vs. OM only.

\section{Effect sizes for proportions}\label{prop1}
Four different effect sizes are common for proportions: risk difference (RD), risk ratio (RR), odds ratio (OR) and number-need-to-treat (NTT) \cite{chowdhury2019}. Because NTT is the inverse of RD  and a specific effect size for randomized clinical trial it will not be discussed here.
The choice between RD, RR, OR is rather complex and not necessarily only determined by design and randomization-therefore they are discussed just as independent approaches. Because proportions are restricted within the range of $[0,1]$, further numerical estimation problems arise which may be quite different for these effect sizes.

\subsection{MCT for odd ratios for proportions}\label{tram}
The canonical link function in the generalized linear model (glm) is the logit link with the log odds ratio as effect size. This will be demonstrated by means of a 2 by k table extracted from the k-by-c table data on quantity of tobacco smoked daily vs. cause of death \cite{chis2} for the categories none and greater1-to-14/day (gr1to14):
\begin{table}[ht]
\centering\footnotesize
\begin{tabular}{rrr}
  \hline
 & None & 1-14/day \\ 
  \hline
Lung\_cancer &  70 & 470 \\ 
  UpperResp\_cancer &   0 & 130 \\ 
  Stomach\_cancer & 410 & 360 \\ 
  Colon-rectum\_cancer & 440 & 540 \\ 
  Prostate\_cancer & 550 & 260 \\ 
  Other\_cancer & 640 & 720 \\ 
  PulmonaryTB &   0 & 160 \\ 
  ChronicBronchitis & 120 & 290 \\ 
  Other\_pulm\_diseases & 690 & 550 \\ 
  Coronary\_thrombosis & 4220 & 4640 \\ 
  Other\_cardiovascular & 2230 & 2150 \\ 
  Cerebral\_hemorrhage & 2010 & 1940 \\ 
  Peptic\_ulcer &   0 & 140 \\ 
  Violence & 420 & 820 \\ 
  Other\_diseases & 1450 & 1810 \\ 
   \hline
\end{tabular}
\end{table}

\small
\begin{verbatim}
library(arm)
library(multcomp)
mod2 <- bayesglm(cbind(gr1to14,none)~cause,data=mydis, family=binomial(link="logit"))
CIOR<-confint(glht(mod2, linfct = mcp(cause = "GrandMean")))
\end{verbatim}
\normalsize

\begin{figure}
	\centering
		\includegraphics[width=0.45\textwidth]{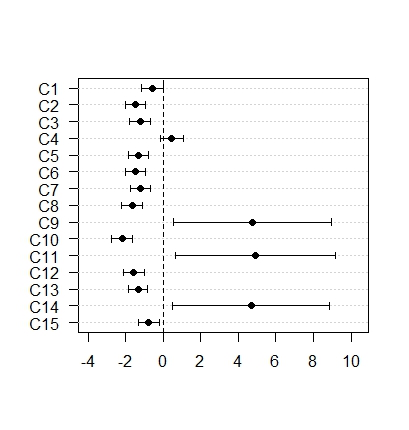}
	\caption{Disease example: log odds ratios}
	\label{fig:logOR}
\end{figure}
There are three significant increase against OM: peptic ulcer; pulmonaryTB, upper respiratory cancer whereas the 15 cause of death are sorted in alphabetic order.

\subsection{MCT for risk differences and risk ratio}\label{RD}
Alternatively, non-canonical links are available that yield risk difference or risk ratio as the effect size. 
Although using library(addreg) to fit additive binomial regression models with identity linkage based on a stable combinatorial EM algorithm is a more general approach \cite{Donoghoe2018}, risk difference as an effect size can also be used by conventional GLM with identity link function under certain conditions. When the proportions in one treatment group, e.g. the control group, are near-to-control, more stable direct estimations instead of plugging-in glm-objects, particularly small $n_i$ approximations, can be recommended \cite{Schaarschmidt2008} available in the package MCPAN.  As an example the  proportions of patients with marked improvement of psoriasis treated with liarozole \cite{berth2000} is used here:

\begin{table}[ht]
\centering\small
\begin{tabular}{rrrrr}
  \hline
 Improvement & Dose150 & Dose50 & Dose75 & Placebo \\ 
  \hline
without &  21 &  27 &  32 &  32 \\ 
  with &  13 &   6 &   4 &   2 \\ 
   \hline
\end{tabular}
\caption{Liarozole example: 2 by k table data}
\end{table}

\begin{figure}[H]
	\centering
		\includegraphics[width=0.450\textwidth]{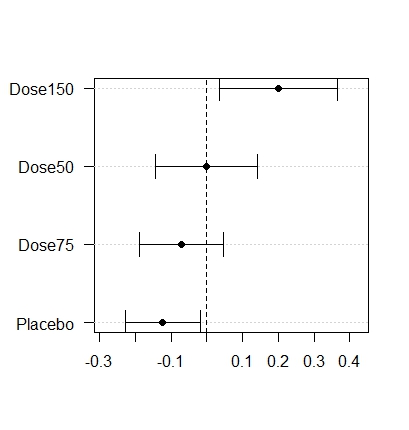}
	\caption{Liarozole example: confidence intervals for differences to OM}
	\label{fig:CIlia}
\end{figure}
From this example it is clear that an ANOVA-type test is not adequate to the problem - here comparisons for control are of primary interest.\\
Risk ratios can be estimated as an effect size from the package \textit{logbin} using an EM algorithm  estimated with the log-binomial model \cite{Donoghoe2018} accordingly.

\section{MCT for hazard ratios for time-to-event endpoints}\label{haz}
For time-to-event data the hazard ratio is the appropriate effect sizes which can be estimated by means of Cox's proportional hazard model, e.g. using the function coxph() in the R-system library survival \cite{surv}.
The data example is part of the Dispenzieri data set \cite{dispenzieri2012} where the relationship of  monoclonal serum immunoglobulin free light chains (FLC) on overall survival in the general population is investigated. Time is defined as days from enrollment until death and event is defined as 0=alive at last contact date or 1=dead. As factor the categorized FLC  levels ($"<50", "50-75", "75-90", ">90"$) for males and females (IA) is considered (cell means model). The Kaplan-Meier-plot demonstrate the survival functions:

\begin{figure}[H]
	\centering
		\includegraphics[width=0.6\textwidth]{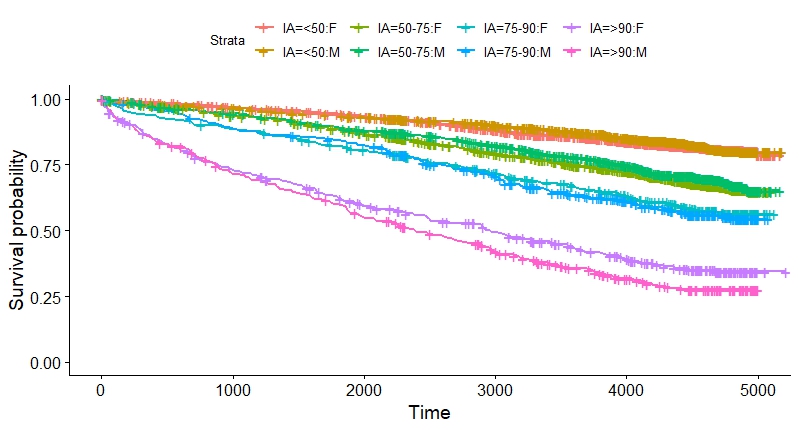}
	\caption{FLC data example: survival functions}
	\label{fig:Surv}
\end{figure}
\small
\begin{verbatim}
library(multcomp)
library(survival)
modfl<-coxph(Surv(futime, death) ~ IA, data =flchain)
simCISU <- confint(glht(modfl, linfct=mcp(IA="GrandMean")))
\end{verbatim}
\normalsize
\begin{figure}[htbp]
	\centering
		\includegraphics[width=0.5\textwidth]{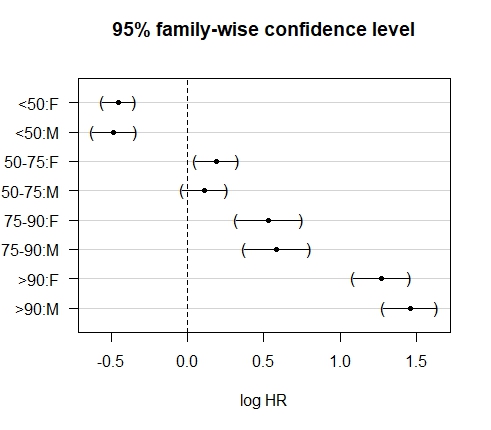}
	\caption{FLC example: simultaneous confidence intervals  for log hazard ratio}
	\label{fig:CIsurv}
\end{figure}

\section{Conclusions}\label{sum}
The alternative approach to the ANOVA F-test presented above is characterized by three main features:
\begin{enumerate}
	\item both a global p-value and local adjusted p-values or compatible simultaneous confidence intervals for comparison with the overall mean are available
	\item different endpoints and models (incl. rank procedures) can be analyzed with an uniform approach
	\item relatively simple realization by means of CRAN packages
\end{enumerate}
Further effect sizes will be considered in future, such as win ratios or Cohen's D, as well as extension to  ANCOVAR and generalized linear mixed effect models.

 \bibliographystyle{plain}
      \footnotesize

\begin{thebibliography}{10}

\bibitem{chis2}
G. Alberti.
\newblock {\em chisquare: Chi-Square and G-Square Test of Independence,
  Residual Analysis, and Measures of Categorical Association}, 2022.
\newblock R package version 0.3.

\bibitem{berth2000}
J~Berth-Jones, G~Todd, PE~Hutchinson, K~Thestrup-Pedersen, and FP~Vanhoutte.
\newblock Treatment of psoriasis with oral liarozole: a dose-ranging study.
\newblock {\em British Journal of Dermatology}, 143(6):1170--1176, 2000.

\bibitem{Brunner2000}
E.~Brunner and U.~Munzel.
\newblock The nonparametric behrens-fisher problem: Asymptotic theory and a
  small-sample approximation.
\newblock {\em Biometrical Journal}, 42(1):17--25, 2000.

\bibitem{chowdhury2019}
S. Chowdhury, R.~C Tiwari, and S. Ghosh.
\newblock Non-inferiority testing for risk ratio, odds ratio and number needed
  to treat in three-arm trial.
\newblock {\em Computational Statistics \& Data Analysis}, 132:70--83, 2019.

\bibitem{Dilba2004}
G.~Dilba, E.~Bretz, V.~Guiard, and L.~A. Hothorn.
\newblock Simultaneous confidence intervals for ratios with applications to the
  comparison of several treatments with a control.
\newblock {\em Methods of Information Ii Medicine}, 43(5):465--469, 2004.

\bibitem{Dilba2007}
G.~Dilba, F.~Schaarschmidt, and L.A. Hothorn.
\newblock Inferences for ratios of normal means.
\newblock {\em R News}, 7:20--23, 2007.

\bibitem{dispenzieri2012}
A. Dispenzieri, J.~A Katzmann, et~al.
\newblock Use of nonclonal serum immunoglobulin free light chains to predict
  overall survival in the general population.
\newblock In {\em Mayo Clinic Proceedings}, volume~87, pages 517--523.
  Elsevier, 2012.

\bibitem{Donoghoe2018}
M.~W. Donoghoe and I.~C. Marschner.
\newblock {logbin: An R Package for Relative Risk Regression Using the
  Log-Binomial Model}.
\newblock {\em {Journal of Statistical Software}}, {86}({9}), {AUG} {2018}.

\bibitem{Hasler2008}
M.~Hasler and L.~A. Hothorn.
\newblock Multiple contrast tests in the presence of heteroscedasticity.
\newblock {\em Biometrical Journal}, 50(5):793--800, October 2008.

\bibitem{Herberich2010}
E.~Herberich, J.~Sikorski, and T.~Hothorn.
\newblock A robust procedure for comparing multiple means under
  heteroscedasticity in unbalanced designs.
\newblock {\em Plos One}, 5(3):e9788, March 2010.

\bibitem{hodges1998}
J. Hodges.
\newblock Some algebra and geometry for hierarchical models, applied to
  diagnostics.
\newblock {\em Journal of the Royal Statistical Society: Series B (Statistical
  Methodology)}, 60(3):497--536, 1998.

\bibitem{Hothorn2008}
L.~A. Hothorn and M.~Hasler.
\newblock Proof of hazard and proof of safety in toxicological studies using
  simultaneous confidence intervals for differences and ratios to control.
\newblock {\em Journal of Biopharmaceutical Statistics}, 18(5):915--933, 2008.

\bibitem{Hothorn2019}
L.A. Hothorn and F.M. Kluxen.
\newblock Robust multiple comparisons against a control group with application
  in toxicology.
\newblock {\em ArXiv:1905.01838 (2019)}, 2019.

\bibitem{hothorn2022IA}
Ludwig~A Hothorn.
\newblock Simultaneous confidence intervals for the interpretation of primary
  and secondary effects in factorial designs without a pre-test on interaction.
\newblock {\em arXiv preprint arXiv:2204.08336}, 2022.

\bibitem{hothorn2023het}
L.~A Hothorn and M. Hasler.
\newblock The dunnett procedure with possibly heterogeneous variances.
\newblock {\em arXiv preprint arXiv:2303.09222}, 2023.

\bibitem{Hothorn2020}
T.~Hothorn.
\newblock Most likely transformations: The mlt package.
\newblock {\em Journal of Statistical Software}, 92(1):1--68, February 2020.

\bibitem{Hothorn2018}
T.~Hothorn, L.~Most, and P.~Buhlmann.
\newblock Most likely transformations.
\newblock {\em Scandinavian Journal of Statistics}, 45(1):110--134, March 2018.

\bibitem{Konietschke2013}
F.~Konietschke, S.~Bosiger, E.~Brunner, and L.~A. Hothorn.
\newblock Are multiple contrast tests superior to the anova?
\newblock {\em International Journal of Biostatistics}, 9(1):63--73, May 2013.

\bibitem{Konietschke2015}
F.~Konietschke, M.~Placzek, F.~Schaarschmidt, and L.~A. Hothorn.
\newblock nparcomp: An r software package for nonparametric multiple
  comparisons and simultaneous confidence intervals.
\newblock {\em Journal of Statistical Software}, 64(9), March 2015.

\bibitem{Lohse2017}
T.~Lohse, D.~Rohrmann, and Hothorn. T.
\newblock Continuous outcome logistic regression for analyzing body mass index
  distributions.
\newblock {\em F1000Res. 2017; 6: 1933.}, 2017.

\bibitem{Pallmann2016}
P.~Pallmann and L.~A. Hothorn.
\newblock Analysis of means: a generalized approach using r.
\newblock {\em Journal of Applied Statistics}, 43(8):1541--1560, June 2016.

\bibitem{Schaarschmidt2008}
F.~Schaarschmidt, M.~Sill, and L.~A. Hothorn.
\newblock Approximate simultaneous confidence intervals for multiple contrasts
  of binomial proportions.
\newblock {\em Biometrical Journal}, 50(5):782--792, October 2008.

\bibitem{segbehoe2022}
L.~S Segbehoe, F. Schaarschmidt, and G.~D Djira.
\newblock Simultaneous confidence intervals for contrasts of quantiles.
\newblock {\em Biometrical Journal}, 64(1):7--19, 2022.

\bibitem{surv}
T.~M Therneau.
\newblock {\em A Package for Survival Analysis in R}, 2022.
\newblock R package version 3.4-0.

\end{thebibliography}

\end{document}